\documentclass{aa}
\usepackage[varg]{txfonts}
\usepackage[utf8]{inputenc}
\usepackage{comment}
\usepackage{amsmath}
\usepackage{graphicx}
\usepackage{gensymb}
\usepackage{upgreek}
\usepackage[colorlinks=true,linkcolor=blue,citecolor=blue]{hyperref}
\usepackage{xspace}
\newcommand{\hei}{\ion{He}{i}\xspace}
\newcommand{\heii}{\ion{He}{ii}\xspace}
\newcommand{\heiii}{\ion{He}{iii}\xspace}

\newcommand{\rbi}{\ion{Rb}{i}\xspace}

\newcommand{\srii}{\ion{Sr}{ii}\xspace}
\newcommand{\yii}{\ion{Y}{ii}\xspace}
\newcommand{\zrii}{\ion{Zr}{ii}\xspace}
\newcommand{\teiii}{\ion{Te}{iii}\xspace}
\newcommand{\Laiii}{\ion{La}{iii}\xspace}

\newcommand{\Ceiii}{\ion{Ce}{iii}\xspace}
\newcommand{\rprocess}{\textit{r}-process\xspace}
\newcommand{\micron}{$\upmu$m\xspace}
\newcommand{\threeS}{1s2s\,$^3$S\xspace}
\newcommand{\threeP}{1s2p\,$^3$P\xspace}
\usepackage{tabularx}

\usepackage{natbib}
\bibpunct{(}{)}{;}{a}{}{,} 

\usepackage{orcidlink}

\begin{document}

\title{Helium features are inconsistent with the spectral evolution \\ of the kilonova AT2017gfo}
\titlerunning{No early helium features in the kilonova AT2017gfo}
\authorrunning{Sneppen et al.}

\author{Albert Sneppen\inst{\ref{addr:DAWN},\ref{addr:jagtvej}}\orcidlink{0000-0002-5460-6126},
Rasmus Damgaard\inst{\ref{addr:DAWN},\ref{addr:jagtvej}}\orcidlink{0009-0002-5765-4601}, 
Darach Watson\inst{\ref{addr:DAWN},\ref{addr:jagtvej}}\orcidlink{0000-0002-4465-8264},
Christine E. Collins\inst{\ref{addr:dublin}}\orcidlink{0000-0002-0313-7817},
Luke~Shingles \inst{\ref{addr:GSI}}\orcidlink{0000-0002-5738-1612} \and
Stuart~A.~Sim\inst{\ref{addr:belfast}}\orcidlink{0000-0002-9774-1192} 
}

\institute{ Cosmic Dawn Center (DAWN)\label{addr:DAWN}
\and
Niels Bohr Institute, University of Copenhagen, Jagtvej 128, DK-2200, Copenhagen N, Denmark\label{addr:jagtvej}
\and
School of Physics, Trinity College Dublin, College Green, Dublin 2, Ireland\label{addr:dublin}
\and
GSI Helmholtzzentrum f\"{u}r Schwerionenforschung, Planckstraße 1, 64291 Darmstadt, Germany \label{addr:GSI}
\and
School of Mathematics and Physics, Astrophysics Research Centre, Queen's University Belfast, Belfast, United Kingdom\label{addr:belfast}
}

\date{Received date /
Accepted date }

\abstract{
    The spectral features observed in kilonovae have revealed the elemental composition and the velocity structures of matter ejected from neutron star mergers. In the spectra of the kilonova AT2017gfo, a P~Cygni line at about \(1\,\mu\)m has been linked to \srii, providing the first direct evidence of freshly synthesised \rprocess material. An alternative interpretation of this feature has been proposed -- \hei\,$\lambda 1083.3$\,nm under certain non-local thermodynamic equilibrium conditions. A key way to robustly discriminate between these identifications, and indeed other proposed identifications, is to analyse the temporal emergence and evolution of the feature. In this analysis, we trace the earliest appearance of the observed feature and detail its spectro-temporal evolution, which we compare with a collisional-radiative model of helium. We show that the \(1\,\mu\)m P~Cygni line is inconsistent with a \hei interpretation both in emergence time and in subsequent spectral evolution. Self-consistent helium masses cannot reproduce the observed feature due to the diminishing strength of radiative pathways out of triplet helium. \newline 
    
    
        }

\maketitle

\section{Introduction}
Observations of the gravitational-wave detected kilonova (KN) GW\,170817/AT2017gfo \citep{Abbott2017b,Coulter2017} revealed the first spectroscopic identification of freshly synthesised \rprocess material in binary neutron star (BNS) mergers \citep{Watson2019}. Notably, studying \rprocess features not only constrains the KN elemental abundances \citep[e.g.][]{Gillanders2022,Vieira2024} but also places detailed constraints on the geometry and velocity structures of the ejecta \citep{Sneppen2023}.

Specifically, the \rprocess species with the strongest known individual transitions \citep[e.g.\ elements from the left side of the periodic table with low-lying energy levels and small partition functions, see][]{Watson2019,Domoto2022} produce the most prominent absorption, emission, or P~Cygni features on the KN continuum in the photospheric epochs. The first of such features to be robustly established, a P~Cygni feature at 700--1200\,nm \citep{Pian2017,Smartt2017}, was interpreted as being due to \srii \ -- which was first identified by \cite{Watson2019} and later corroborated with additional radiative transfer modelling \citep{Domoto2022,Gillanders2022,Vieira2023,Shingles2023} and in non-local thermodynamic equilibrium (NLTE) models \citep{Pognan2023}. The strontium mass required to produce the feature is compatible with that produced in the dynamical ejecta from merger simulations \citep{Perego2022}. Independent evidence for the \srii identification is that strong UV absorption is observed at $\lambda \leq400$\,nm around 1.4\,days after merger, which could be explained by lines from co-produced elements, including \yii \citep{Gillanders2022} and \yii and \zrii \citep{Vieira2023}. Further corroboration of the \srii interpretation comes from the existence of a 760\,nm P~Cygni feature, interpreted as being due to multiple lines from \yii \citep{Sneppen2023b} and the timing of reverberation recombination-waves moving across the ejecta \citep{Sneppen2024}. For completeness, we also note that second \rprocess peak elements have been suggested (viz.\ \Laiii and \Ceiii) in order to explain near-infrared (NIR) features at intermediate times \citep{Domoto2022}. It is worth noting that the observed spectral features are from \rprocess species and can be well explained by assuming LTE conditions at the observed black-body temperature \citep{Sneppen2024}. However, non-thermal particles produced in the radioactive decay of the \rprocess nucleosynthesis could allow for ionisation states and energy levels outside the LTE limit.

\begin{figure}
    \centering
    \includegraphics[width=\linewidth,viewport=24 20 590 608 ,clip=]{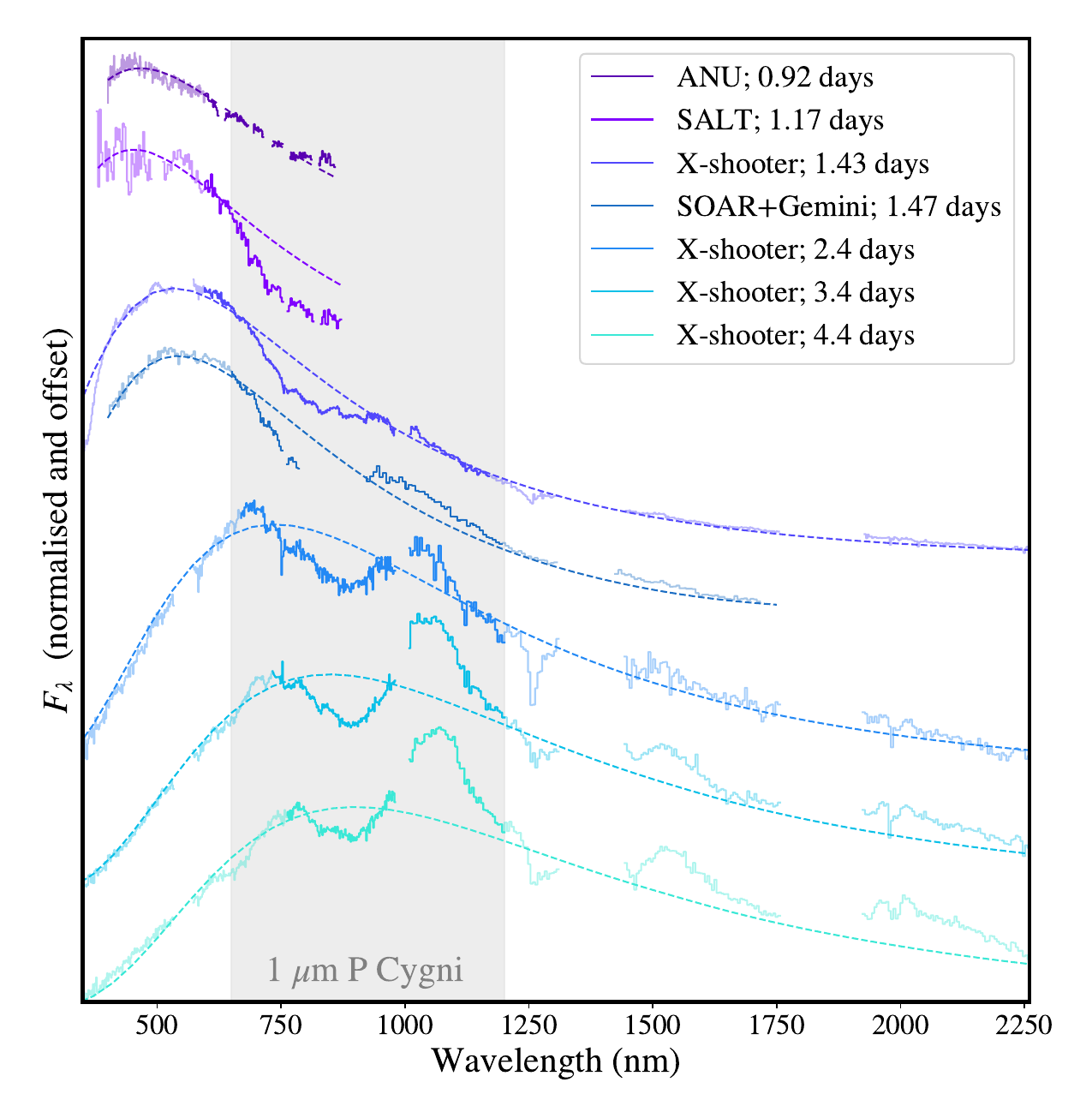} 
    \caption{Evolution of the 1\,$\mu$m P~Cygni feature over the first five days post merger from the unified spectral compilation \citep{Sneppen2024}. The spectra were taken post merger plus 0.92\,day \citep[ANU,][]{Andreoni2017}, 1.17\,day \citep[SALT,][]{Buckley2018}, 1.47\,day \citep[SOAR and Gemini-south,][]{Nicholl2017,Chornock2017} 1.43, 2.4, 3.4, and 4.4\,day \citep[X-shooter,][]{Pian2017,Smartt2017}. The spectra prior to one day contain no strong deviation from the black-body continuum \citep[see also][]{Shappee2017}. However, between 0.92 and 1.17\,days, strong absorption appears, while an emission feature begins to emerge with a delay of around ten hours after this (1.43-1.47\,days). We note that these are the predicted emergence time framess for both absorption and emission for a feature produced by LTE \srii due to a recombination wave passing through the ejecta \citep{Sneppen2023_rapid}. We show in this analysis that this emergence is inconsistent with a \hei\,$\lambda 1083.0$\,nm interpretation of the feature.} 
    \label{fig:emergence_of_feature}
\end{figure}

Indeed, an alternative interpretation for the 1.0\,\micron P~Cygni feature has been proposed, namely that it could originate from the neutral He\,{\sc i}\,$\lambda$1083.3\,nm line (\threeS-\threeP transition), which in certain NLTE conditions can produce a feature at a similar wavelength \citep[see discussions in][]{Perego2022,Tarumi2023}, while \srii may potentially be ionised away by a population of non-thermal electrons from radioactive heating. As the spectrum becomes nebular towards later epochs, after \(t \gtrsim 5\)\,days post merger, NLTE ionisation modelling should become important \citep{Pognan2022}, and several features (which would be negligible in LTE conditions) have been suggested for these epochs, including \rbi \citep{Pognan2023} and \teiii \citep{Gillanders2023,Hotokezaka2023}. Nevertheless, the validity of LTE, alongside the transition point to when NLTE modelling is required, has so far been difficult to determine. 
In this context, helium is a particularly interesting species to consider, as it requires NLTE conditions and the required atomic data is largely known and has been applied in many supernova studies \citep[e.g.][]{Lucy1991}.

In this paper, we therefore revisit NLTE helium proposition to determine if the observed feature can be self-consistently modelled by \hei under NLTE in the early epochs of AT2017gfo (from 0.92--4.4\,days post merger). In Sect.~\ref{sec:observed_emergence}, we briefly summarise the observed evolution of the observed 1\,\micron feature in the early epochs. In Sect.~\ref{sec:he}, we discuss the assumptions required to produce an observable He\,{\sc i}\,$\lambda$1083.3\,nm feature. In Sect.~\ref{sec:he_NLTE_calculation}, we enumerate the various relevant transitions between states and implement a collisional-radiative model for helium. 
In Sect.~\ref{sec:he_evolution}, we summarise the predicted evolution of a helium feature and show that the modelled evolution is inconsistent with AT2017gfo's spectral series. 
Lastly, in Sect.~\ref{sec:NLTE_required}, we discuss the broader implications this reveals on the radioactive ionisation and the validity of LTE ionisation in the ejecta.

\section{The observed evolution of the 1\,\micron feature}\label{sec:observed_emergence}
To provide the strongest constraints for modelling the 1\,\micron feature, we examine the first appearance and subsequent evolution of the feature. In \cite{Sneppen2024}, we have compiled the early spectra of AT2017gfo from Magellan, ANU, SALT, VLT/X-shooter, and Gemini in order to provide a higher cadence analysis at early times than any individual data series provides (see Fig.~\ref{fig:emergence_of_feature}). 
This revealed that the first detected appearance of the 1\,\micron P~Cygni feature is from SALT \citep[1.17\,days post merger,][]{Buckley2018}, with an absorption component at around 800\,nm. The spectrum taken six hours earlier from ANU \citep[0.92\,days post merger,][]{Andreoni2017} displays no clear deviation from an underlying black-body continuum. Notably, the flux calibration of the ANU spectral energy distribution (SED) is somewhat uncertain, so the exact black-body temperature is not strongly constrained from the spectral slope \citep{Sneppen2024}. However, this does not strongly affect the non-detection of a 1\micron feature in absorption, as this would generate a distinct spectral break in the middle of the spectrograph (similar to the SALT spectrum).
Thus, the feature emerges rapidly over a relatively short timescale of a few hours. From 1.4\,days post merger, AT2017gfo was monitored at a nightly cadence with the X-shooter spectrograph mounted on the European Southern Observatory's Very Large Telescope \citep{Pian2017,Smartt2017}. The spectral range (330--2250\,nm) has provided the first broad spectral coverage encompassing the near-infrared (NIR) emission peak of the 1\,\micron P~Cygni feature. At 1.43\,days, this feature had not yet formed in emission, but over the subsequent hour the feature formed in emission as well as absorption producing a full P~Cygni profile \citep{Sneppen2024}. If the feature was due to strontium, this would be due to light travel time delay and reverberation effects, where the more distant ejecta (where less time has passed post merger) has not yet cooled sufficiently to recombine with \srii \citep{Sneppen2023_rapid}. In contrast, as we show in this analysis, a helium feature will not produce such a rapid emergence and therefore not be sensitive to reverberation effects to the same extent. Towards the later spectra, the characteristic velocities decrease as the photosphere recedes deeper into the ejecta, and the optical depth of the absorption decreases. Post five days, the feature becomes more emission dominated -- a result that would be produced by reverberation effects \citep[see][]{Sneppen2023_rapid} but that may also have a contribution from the decreasing densities reducing the optical depth in the line(s) and allowing radiative de-excitation to form pure emission features \citep[see][]{Gillanders2023}. In summary, the observed 1\micron feature is not detected at 0.92\,days, indicating it is optically thin. It is marginally optically thick from 1.17-day to 3.4-day spectra and then becomes (increasingly) optically thin from the 4.4-day spectrum and onwards. 


The underlying continuum is well described as a Planck function, $B(\lambda,T)$, with the temperature, \(T\), at wavelength \(\lambda\) emitting from a sharply defined photosphere. At early times, a black body provides excellent fits to the data with percent-level consistency in the required temperature from the UV through the NIR \citep{Sneppen2023_bb}. Towards the later photospheric epochs (see Fig.~\ref{fig:emergence_of_feature}), NIR emission features emerge at 1.6 and 2.0\micron, which if not modelled will bias a simple best-fit black body towards lower temperatures than inferred from the UV side. As the object is dramatically cooling in early epochs, the observed feature persists over a broad range of temperatures from 4900-2900\,K (1.17-3.4 days), which over the subsequent days then fades away, while the temperature cools a relatively minor few hundred Kelvin. 


\section{He\,{\sc i}\,$\lambda$1083.3\,nm line}\label{sec:he}

\begin{figure}
    \includegraphics[width=\columnwidth,viewport=15 15 418 384,clip=]{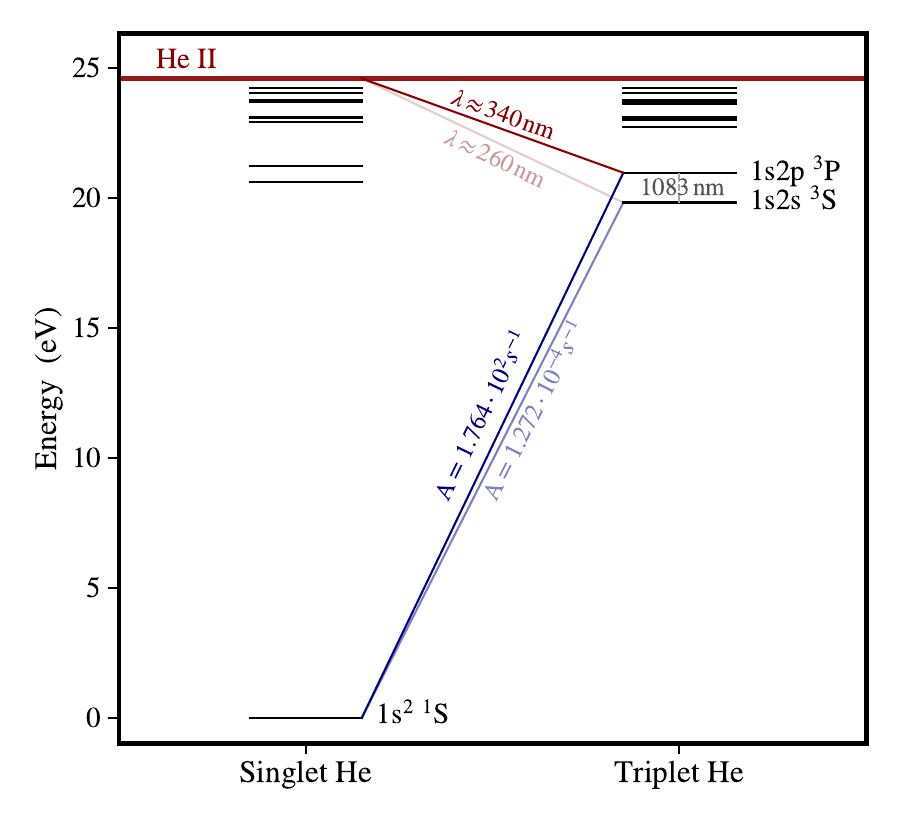}
    \caption{Grotrian diagram showing the \hei energy levels. The \threeS--\threeP 1083\,nm transition could produce a 1\,\micron P~Cygni feature in NLTE models \citep[e.g.][]{Tarumi2023} because \threeS can be difficult to ionise (e.g.\ requiring UV photons $<260\,$nm) and the natural transition to the ground state is very slow (e.g.\ characteristic timescale of hours). However, as shown here, the other well-populated triplet state, \threeP, is several orders of magnitude more susceptible to ionisation and naturally decays to the ground state \(10^6\) times faster, on a timescale of 0.01\,second.} 
    \label{fig:grotrian}
\end{figure}

The critical aspect in whether He\,{\sc i}\,$\lambda$1083.3\,nm (\threeS--\threeP) can explain the observed feature relies on the population of the lower energy level of the transition, which is the lowest-lying level (pseudo ground state) of the triplet states of neutral helium (see Fig.~\ref{fig:grotrian}). There are three key conditions required to produce a \hei feature:

\begin{enumerate}
    \item Naturally, one needs an appropriate abundance of helium, which in contrast to most \rprocess elements varies by as much as four orders of magnitude between different nucleosynthesis simulations. Minor helium abundances can be produced when modelling dynamical ejecta \citep[such as][where $X_{\rm He}\approx0.01\%$]{Perego2022}, while high helium abundances can be produced for ejecta with high $Y_e$ \citep[such as][where $X_{\rm He}\approx1-50\%$]{Kullmann2022,Kawaguchi2022,Just2023}. \vspace{3pt}
    \item As the triplet ground state is 19.8\,eV above the \hei ground state (compared with $k_B T \approx 0.25$--$0.4$\,eV for early epochs in AT2017gfo), the triplet population is not efficiently populated by excitation due to thermal particles. The triplet states are instead populated by recombination from \ion{He}{ii}, which is itself populated by ionisation of \hei by radioactive decay particles. This means that most effective populating of the \hei triplet ground state occurs when the majority of helium is singly ionised.\vspace{3pt} 
    \item The triplet levels are also highly sensitive to photoionisation from near-UV photons due to their proximity to the ionisation energy (\(\lesssim5\)\,eV). The most detailed NLTE helium modelling in the KN context up to this point \citep{Tarumi2023} only includes ionisation from the triplet ground state, \threeS (e.g.\ photons with energies $>4.8$\,eV, $\lambda<260$\,nm). However, in our analysis, we find that ionisation from the \threeP-state ($\lambda\approx340$\,nm) must be included because the photoionisation and natural decay pathway leaving the triplet states in all early epochs is not dominated by the \threeS state but by the \threeP state, as we discuss in Sec.~\ref{sec:helium_sensitivity}. 
\end{enumerate}

We focus in this modelling on the last issue because this allows for quantitative predictions to be made for the temporal evolution of the \hei feature. The expected evolution can be compared and contrasted with the expected evolution of features due to strontium and with the 1\,\micron feature as it is observed. In particular, this aspect is most strongly constrained observationally in the earliest appearance of the feature when comparing with the UV flux constraints from \textit{Swift}-UVOT, which we turn to next.

\begin{figure}
    \includegraphics[width=\columnwidth,viewport=18 16 493 417,clip=]{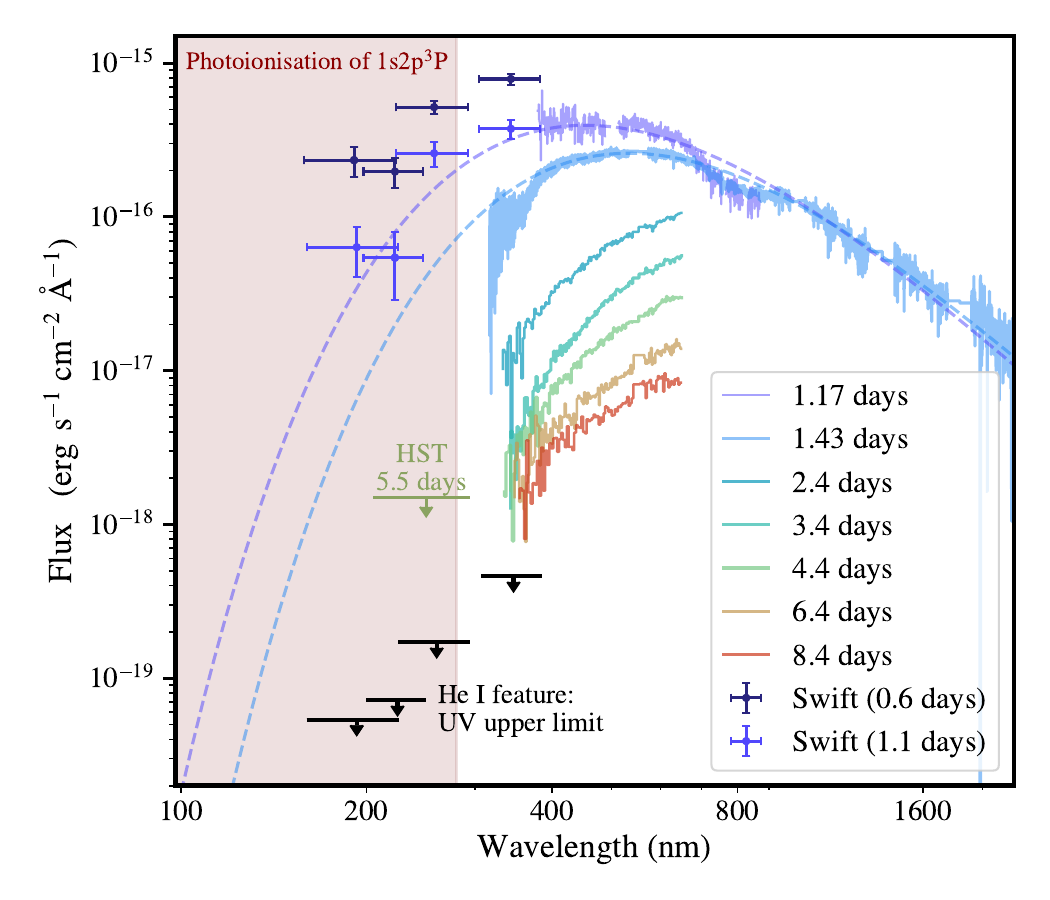} 
    \caption{SALT and VLT/X-shooter spectra with \emph{Swift}-UVOT photometric constraints on the early UV flux. Photons with energies sufficient to ionise the population at 1s2s\,$^3$P ($\lambda < 340$\,nm rest frame or equivalently $\lambda < 280$\,nm corrected for ejecta expanding at $0.28c$) is shown by the red region. The UV flux detected by \emph{Swift}-UVOT in \cite{Evans2017} decreases with time, falling below the 3$\sigma$ (2$\sigma$) detection threshold after 1.1 days (3.0 days). For comparison with the \emph{Swift} observations, the maximum UV flux that allows \hei to still produce the 1\,\micron feature (within the \citet{Tarumi2023} NLTE helium model) is shown in black. The \emph{Swift} flux is around three orders of magnitude larger than what is permitted for the helium interpretation.} 
    \label{fig:swift}
\end{figure}


The relevant energy levels (1s2s$^3$S at 19.8\,eV and 1s2p$^3$P at 21.0\,eV above the ground state) are sensitive to photoionisation to \heii (at 24.6\,eV above the ground state) because UV photons (with $\lambda < 260$\,nm for \threeS and $\lambda < 340$\,nm for \threeP) would quickly depopulate these triplet states of helium through photoionisation. \citet{Tarumi2023} note that 0.3\% of the flux inferred from the Wien tail of a 4400\,K black body (e.g.\ approximately the Doppler-corrected 1.4-day black body) is sufficient to depopulate the triplet states in their models, but given the lack of UV observations, it was tentatively suggested that line blanketing could suppress the UV flux to this level (and thus permit an NLTE population in the \threeS state). 

However, nearly contemporaneous with the SALT spectrum (and the first chronological observation of the 1\,\micron feature), there are observational constraints on the UV flux taken with \emph{Swift}-UVOT \citep{Evans2017,Nicholl2017}. We show these constraints in Fig.~\ref{fig:swift}, which suggest i) the \emph{Swift} data naturally extends the optical black body into the UV (showing no wholesale UV blanketing) and ii) the photometric observations are far in excess of the model-allowed UV flux upper limit, which would thus photoionise and hence depopulate the triplet states. Given the observed decrease of the UV flux from 0.6 to 1.1\,days (and that we are three orders of magnitude above the \cite{Tarumi2023} UV upper limit), the UV flux is likely to have been too large to permit a sufficient \hei triplet population to form a strong feature -- thus excluding the \hei interpretation -- in the first days post merger. In the subsequent days, a lower UV flux was tentatively detected -- with the last detection (at $2\sigma$ significance) at 3.0\,days post merger in the \emph{Swift} UVW2 band. While extrapolating the \emph{Swift} data beyond three days would be quite uncertain, the consistency of the UVOT photometry with a single temperature black body inferred from the X-shooter data in the first few days with no significant evidence for line blanketing gives some confidence that extrapolating the UV flux in wavelength from black-body fits to the X-shooter UVB arm at four to five days may be reasonable. Such spectral extrapolation strongly suggests the UV flux still exceeds the upper limit until at least four to five days post merger. An \emph{HST} STIS UV spectrum provides a 3$\sigma$ upper limit of \(F_{UV} \lessapprox 1.5\times10^{-18}\)\,erg\,s\(^{-1}\)\,cm\(^{-2}\)\,\AA\(^{-1}\) at 250\,nm at 5.5\,days \citep{Nicholl2017}, which ultimately is not below the limits of the UV flux expected from extrapolations of the X-shooter SED. 


\begin{figure*}
    \includegraphics[width=\textwidth]{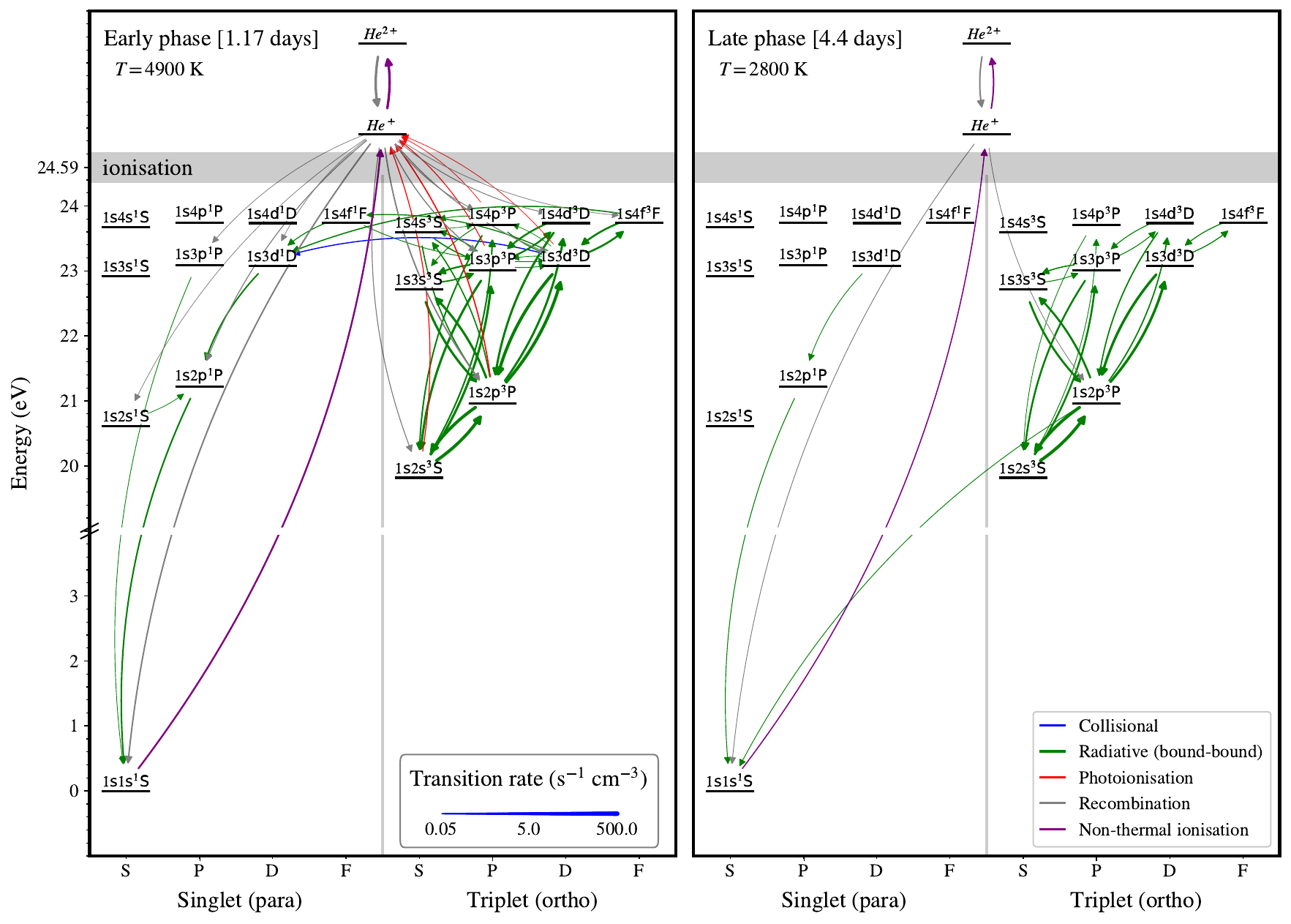}  
    \caption{Network Grotrian diagram indicating the dominant pathways entering and leaving any level. The arrow-width shows the transition rate on a logarithmic scale, with the colour indicating the dominant mechanism. At early times, the dominant pathways leaving triplet \hei is photoionisation. At later times (as the upper triplet levels are less populated), natural decay to the \hei ground state and collisional pathways out of the triplet become increasingly important. }%
    \label{fig:Networks_diagram}
\end{figure*}


\section{ Collisional-radiative NLTE model }\label{sec:he_NLTE_calculation}

To accurately model helium under NLTE at various epochs, we have implemented a collisional-radiative model. In Fig.~\ref{fig:Networks_diagram}, we show an energy-level diagram and highlight the various pathways and rates connecting the \hei levels. To produce a strong \hei\,$\lambda 1083.3$\,nm (\threeS-\threeP transition) feature requires the triplet (orthohelium) population i) to be decoupled from the singlet (parahelium) states, as these rapidly radiatively de-excite to the (isolated) ground state, and ii) to have only weak pathways for photoionisation to \heii. Crucially, accurately modelling the \threeS level population must include higher energy levels and their transitions, as these contribute significantly to the pathways depopulating triplet helium. We have collisional data between all levels with $n \leq 4$, so this constitutes the fiducial model in this work. We find that including further energy levels is of lesser importance given their small NLTE populations, with less than a 10\% decrease in the triplet population when including all levels up to $n \leq 8$ for all epochs. 

\subsection{The transition channels}
In the following, we enumerate the transition channels between levels. The \citet{Tarumi2023} model includes collisional de-excitation, bound-bound radiative excitation, and photoionisation directly leaving the \threeS state. The difference between the work of \citet{Tarumi2023} and the model presented here is that we include all the direct and indirect pathways that connect triplet and singlet states, that is, the inclusion of collisional, radiative, and photoionisation pathways from the triplet states above \threeS. As \threeS is strongly coupled to these higher triplet levels, the pathways from higher energy levels are important (see Fig.~\ref{fig:Networks_diagram}).


In our model, for ease of comparison, we assume an identical ejecta structure to \citet{Tarumi2023}, namely, a spherically symmetric and homogeneous ejecta with a mass of $0.04M_\odot$, distributed as \(\rho=\rho_0 v^{-5}t^{-3}\) between $v=0.1c$ and $v=0.5c$, with the value of $\rho_0$ set by the total helium mass in the ejecta. The electron density has the same spatial distribution as \(\rho\), as the free electrons trace the lowly ionised \rprocess elements -- for the electron distribution we follow \citet{Tarumi2023} and assume \(n_e=1.5\times10^8\,\)cm\(^{-3}\) (at $v=0.3c$, $t=1$\,day) as our default model. We have also explored using other ejecta structures with the ultimate conclusions on the self-consistency of the model remaining unchanged (see Sect.~\ref{sec:variation_ejecta_structure}). 


The line-forming region is illuminated by a black-body photon background from the underlying photosphere. Specifically, we assumed a geometrically diluted black-body radiation field inferred from the Doppler-corrected continuum. In practice, this means the best-fit black body to the continuum is assumed to set the local radiation field (needed for modelling radiative transitions) and the electron temperature (needed for modelling recombination and collisional transitions). Notably, this uses a photospheric approximation over the various wavelengths of scattering, which requires the assumption that the radiation field inferred (e.g the Doppler-corrected black body) is representative of the local radiation field. Specifically, the outer line-forming region $\sim0.45c$ ($\sim0.30c$) scatters 10-20\% lower wavelength than that expected from the inner line-forming region at $\sim0.3c$ ($\sim0.2c$) at 1.17\,days (4.4\,days). The continuum from such different wavelengths could originate from different velocity layers (characterised by different Doppler-corrections, temperatures, and radiation fields). Observationally, the continuum remains well modelled as a black body emanating from a single photosphere in the first days post merger, but employing a photospheric approximation from UV through NIR is tenuous. Regardless, the assumption employed here is only over a relatively limited range in wavelength of a few hundred nanometers.
Lastly, the geometric dilution \(W(v) = 0.5 \cdot \left(1-\sqrt{1-(v_{ph}/v)^2}\right) \) is included to represent that the radiation field decreases for velocity shells that are increasingly distant from the photosphere \citep[e.g.][]{Mihalas1978}.  

\subsubsection{Radiative transitions: Bound-bound}
    At all times investigated, radiative transitions within triplet levels dominate over all other processes and thus set the relative level population. We modelled the three different types of radiative transitions, namely, absorption and spontaneous and stimulated emission. 
    
    We used $A$ values for all transitions extracted from NIST \citep{Kramida2023}. For multiple lines between different fine-structure levels, we summed the transition's lower levels and averaged over all upper states involved \citep{Axner2004}. Notably, the transition strengths between triplet and singlet states systematically increase for higher energy levels. In particular, the 1s4d$^3$D--1s4f$^1$F and 1s4f$^3$F--1s4d$^1$D transitions are very strong, $A\sim10^8\,$s$^{-1}$ \citep{Drake2006}. We used the Sobolov escape probability, \(P_{esc}(\tau) = (1-e^{-\tau})/\tau \), and the blue-wing intensity from the black-body model to describe the mean intensity in each line \citep[e.g following the procedure for homologous fast velocity fields in][]{Lucy2002}. 
    
    We have not included any additional radiative decay mechanisms, such as two-photon emission, since these occur on much longer timescales. In the limit of detailed balance between two levels, the level population ratios are Boltzmann-distributed with a geometric dilution factor of 0.5 (as exemplified in Eq.~\ref{eq:triplet}, below). Due to the strong transitions between triplet states, this provides a good approximation of the relative triplet helium populations at all times investigated.

    \subsubsection{Collisional transitions} 
    In the low-temperature limit, transitions from triplet to singlet states are driven by collisions with thermal electrons. We used the thermally averaged transition rates from \citep{Ralchenko2008}. We are not particularly sensitive to the exact transition rates employed with the collisional transition rates from \citet{Berrington1987} yielding similar conclusions. 
    The rate of ionisation by thermal electrons is negligible compared to the collisional de-excitation rate, which is similar to the findings in previous works \citep[e.g.][]{Tarumi2023}.

    \subsubsection{Photoionisation}
    The wavelength-dependent photoionisation cross-section, $\sigma_{\rm PI}$, for different levels is taken from \citet{Nahar2010}.
    In the first days post merger, photoionisation is a significant source of depopulation. 
    For higher levels, the photoionisation cross-section increases, and more of the spectral flux can ionise as the threshold photon energy decreases. 
    The \threeS level has the smallest photoionisation cross-section of any triplet species and only extends up to 260\,nm. \threeP level has a photoionisation cross-section that is approximately five times greater and extends up to 340\,nm. The levels from $n=3$ and above can be photoionised from photons near the black-body peak, where the photon density is at its highest. This means that photoionisation of $n=3$ levels is unaffected by UV flux modelling. While \threeS will have a larger population than higher energy levels, this level is never a dominant photoionisation pathway for any observed KN temperature.
    Photoionisation does not affect \heii around or above the photosphere (given the black-body radiation field observed) due to its high ionisation potential of $53$\,eV.

    \subsubsection{Recombination}
    For recombination, we used the temperature-dependent state-specific recombination coefficients, $\alpha$, provided in \citet{Nahar2010}. The recombination flux was thus modelled on a level-by-level basis, but due to the natural transition cascades from excited levels, the excited electron quickly decays to the triplet and singlet ground states. For all relevant temperatures, this implies that approximately half of the recombinations effectively end up populating 1s$^{2}$\,$^1$S and 1s2s\,$^1$S, while the remaining half populates \threeS. 


    \subsubsection{Non-thermal electron ionisation}\label{sec:hot_electron}
    Nuclear decay creates non-thermal electrons that can ionise helium, thus producing the \heii required for recombination to populate triplet helium. 
    We modelled this process in a manner identical to \citet{Tarumi2023} and did not revisited their approximations for evaluating the Spencer-Fano calculations. Specifically for ionisation, we assumed a non-thermal electron flux from $\beta$-decay of \rprocess elements from \citet{Hotokezaka2020} of $1$\,eV\,s$^{-1}t^{-1.3}$ per ion,
    with $t$ in days and a work per ion of 593\,eV for \hei and 3076\,eV for \heii (from \citealt{Tarumi2023}, who refer to work in preparation by Hotokezaka et~al.). 
    Non-thermal electrons can ionise any level of \hei, but this process is only the limiting rate step for the ground state. For all energy levels above the ground state, the radiative and thermal collisional pathways dominate. While we model this process identically for every \hei level, it is only important from the ground state. For helium, the main uncertainty in non-thermal ionisation lies in the deposition rate, as (to the first order) the ionisation rate coefficients scale linearly with the deposition rate. The deposition rates are similar across models within a factor of a few, which implies the dominant uncertainty of the helium ionisation state comes from the recombination rate (i.e. the exact electron density).

    It is worth noting that the non-thermal modelling leads to the most generous conditions for the triplet helium population. At all times considered, \heii is a dominant species near the photosphere, which maximises the recombination rate and thus the possible triplet density. If one were to significantly increase or decrease the radioactive ionisation, \heii could become sub-dominant to either \hei or \heiii.


\subsection{Inferred He mass}
Using this NLTE population modelling, we were able to estimate and compare the helium mass needed to produce the observed feature. Specifically, the NLTE model determines what fraction of helium is in the \threeS level for any given epoch, where we infer the radiation field (needed for radiative transition) and the particle temperature (needed for collision rates and recombination rates) from the Doppler-corrected observed black-body continuum in each epoch. Given a total helium density and the time post merger, the NLTE model thus sets the \threeS density across the line-forming region. To produce a Sobolev optical depth, $\tau$, given the oscillator strength of the \hei\,$\lambda 1083.3$\,nm feature requires a density n(\threeS)\(\approx  7.4 \tau t_d^{-1}\,\)cm\(^{-3}\) where $t_d$ is the time in days \citep{Tarumi2023}. 

The Sobolev optical depth (at the photosphere) is inferred from a best-fit P~Cygni to the absorption feature, where we have used the P~Cygni implementation in the Elementary Supernova model \citep[see][]{Jeffery1990}.\footnote{We adapted Ulrich Noebauer's \texttt{pcygni\_profile.py} in \url{https://github.com/unoebauer/public-astro-tools}} Here the profile is expressed in terms of the rest wavelength $\lambda_0$ and the line optical depth $\tau(v) = \tau(v_{\rm ph}) \cdot e^{-(v-v_{\rm ph})/v_{\rm e}}$, with a scaling velocity $v_{\rm e}$, a photospheric velocity $v_{\rm ph}$, and a maximum ejecta velocity $v_{\rm max}$. Nearly all of these parameters can be optimised to fit the data, with the exception of the known rest wavelength of the transition, ie. $1083.0$\,nm. We note this central wavelength is slightly redward (0.02-0.06c) of the observed emission peak in all photospheric epochs. The parameters related to velocity can be chosen to mimic the observed feature velocity scale as the ejecta-structure is a priori (largely) unknown. The optical depth of a line sets the \threeS density or, by extension (given the collisional-radiative NLTE model), the total helium density. From this, we could deduce the total mass above the photosphere by integrating over each velocity shell 
\begin{equation}
    M_{\rm He} = \int 4\pi (v t)^2 \ \rho X_{\rm He} \ d(vt)\label{eq:mass} 
\end{equation}
from the inner photospheric shell and outwards. For the baseline model, we assumed $X_{\rm He}$ is independent of velocity, but for completeness we did examine the potential impact of composition gradients in Sect.~\ref{sec:he_evolution}. Given the rapid decline of density with velocity, \(\rho \propto v^{-5}\), our mass is dominated by the ejecta near the photosphere. The resulting masses are detailed in Fig.~\ref{fig:he_mass_time}. Within this framework, the dominant uncertainty in the required helium mass comes from the exact \threeS level-population relative to the total helium population, which is dominated by the uncertainty in the ejecta temperature. We emphasise this `helium mass' is tightly coupled to solely the \heii mass, as this is a dominant ion. 

In Fig.~\ref{fig:comparing_emergence}, we show how the He\,{\sc i}\,$\lambda$1083.3\,nm feature would evolve across epochs given various helium mass fractions. The continuum is modelled by the best-fit black-body overlayed with the P~Cygni parameterised from the elementary supernova model (deliberated in Sect.~\ref{sec:observed_emergence}), which has an optical depth set by the \threeS density. The reference total ejecta mass of the high-velocity ejecta ($v\gtrapprox0.2c$) is around 0.01-0.02\,M$_\odot$ \citep[e.g.\,][]{Cowperthwaite2017,Villar2017,Siegel2019}. It is likely that there is even less mass in the line-forming region from 0.92 to 2.4\,days because the velocity $v\gtrapprox0.25-0.3c$ \citep[inferred from the 1\micron line or the black-body continuum, see detailed velocity-constraints in][]{Sneppen2023}, but as the photosphere recedes towards $0.2c$, this estimate likely becomes representative of the total ejecta mass around three to five days post merger.

\subsection{NLTE helium modelling: Sensitivity to \threeP}\label{sec:helium_sensitivity}
\citet{Tarumi2023} concluded that a \hei feature could be relatively stable when examining the pathways leaving \threeS in the limit of no UV flux. When including all the relevant energy levels, as we show below, a \hei feature's strength is however temperature sensitive and would evolve by orders of magnitude between epochs. As we present in Sect.~\ref{sec:he_evolution}, this implies that a \hei feature cannot match the evolution observed in AT2017gfo over the photospheric epochs. 

The presented NLTE modelling of helium, assuming a geometrically diluted black-body radiation field, indicates that the relative level populations are well described as Boltzmann distributed with a dilution factor of $W\approx0.5$ for \threeP \citep[see][] {Bhatia1986,Tarumi2023}, 
\begin{equation}
    \frac{n({\rm 1s2p\,^3P}) }{n({\rm 1s2s\,^3S})} = \frac{3}{1} {\rm exp} \left[ \frac{E({\rm 1s2p\,^3P})-E({\rm 1s2s\,^3S})}{k_BT} \right] W.
    \label{eq:triplet}
\end{equation}
For the de-reddened black-body temperatures $T$ for the early 1\,\micron feature (i.e.\ 5000\,K and 4200\,K at 1.17 and 1.43\,days), this would imply 10\% and 6\%, respectively, of the triplet helium is in the \threeP level. 
We again note that in this model we equate the temperature of the local black-body radiation field with the observed Doppler-corrected black-body temperature. 
However, even perturbing this assumption, we are almost certainly in the regime where a few percent of the population is in higher triplet levels.

This subtle minority of the triplet population is important in two respects. Firstly, the photoionisation cross-section for \threeP is approximately five times larger than for \threeS \citep{Nahar2010}, and it can photoionise with redder photons (i.e.\ $\lambda\approx340$\,nm), where the flux level is at least two to 20 times larger, as it is closer to the spectral peak. For the relevant temperatures of AT2017gfo where the 1\,\micron feature is present, the photoionisation rate from \threeP is around one to two orders of magnitude larger than from \threeS. \threeS and \threeP are photoionised respectively within $\sim0.01$\,s and $\sim0.0005$\,s by 1.17 days and $\sim300$\,s and $\sim1$\,s by 4.4 days. Secondly, the natural decay to the ground level from \threeP (having a lifetime of $\sim0.01$\,s) is a million times faster than \threeS (which has a lifetime of roughly two hours). It is thus only at late times that natural decay becomes important for estimating the triplet population. 

Including even higher triplet energy levels (with quantum numbers n=3, n=4, etc.) adds to the depopulation channels but only significantly at early times. This is because these higher energy levels are very weakly populated in both LTE and NLTE ($\approx 10^{-4}$ less populated than \threeS at one to two days post merger and rapidly dropping with time). However, the depopulation rate from these higher states due to photoionisation is also $O(10^4)$ times larger, as optical photons near the spectral peak can ionise, and the cross-section for photoionisation is larger. Additionally, some of their radiative decays to the singlet states are very large, with $A$ values on the order of $10^6\,$s$^{-1}$. Thus, these higher levels are actually comparable in importance to \threeP at early times.



\section{Evolution of a NLTE \hei feature}\label{sec:he_evolution}

\begin{figure}
    \includegraphics[width=0.96\linewidth,viewport=14 16 452 450 ,clip=]{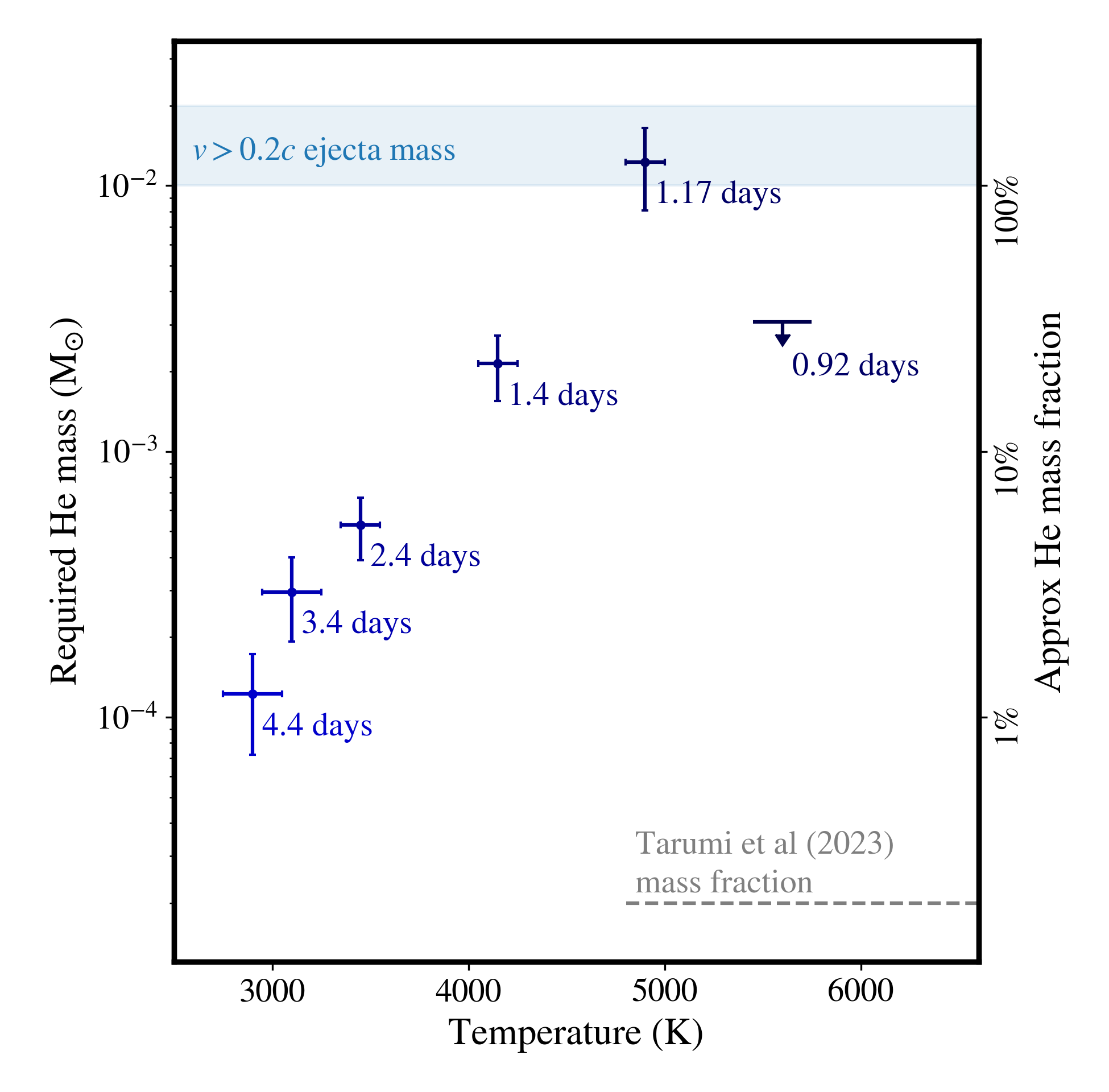}
    \caption{Estimated helium mass required to produce the absorption of the 1\,\micron feature in early spectra. These estimates are derived using the NLTE helium model described in Sec.~\ref{sec:he_NLTE_calculation}. 
    The left y-axis indicates the approximate He mass fraction (given the high velocity ejecta component is around $\sim$0.01\,M$_\sun$, which is likely most representative around three to five days post merger), while the exact helium mass fraction for each epoch can be found in Table~\ref{table:table}.
    The required helium mass changes drastically between epochs, and for any epoch, it is largely inconsistent with all remaining epochs (primarily due to the decreasing strength of radiative pathways leaving triplet helium). 
    At early times, the helium mass required for a feature is comparable to the total mass of the early  higher-velocity ($v\approx0.2$-$0.3c$) `blue' ejecta. The absence of a helium feature at 0.92\,days provides a $5\sigma$ upper limit on the helium mass (reported as solely the statistical uncertainty from the observed spectra). The grey dashed line indicates the mass fraction considered in \cite{Tarumi2023}.}
    \label{fig:he_mass_time}
\end{figure}

\begin{figure}
    \includegraphics[width=0.5\textwidth,viewport=18 15 360 512, clip=]{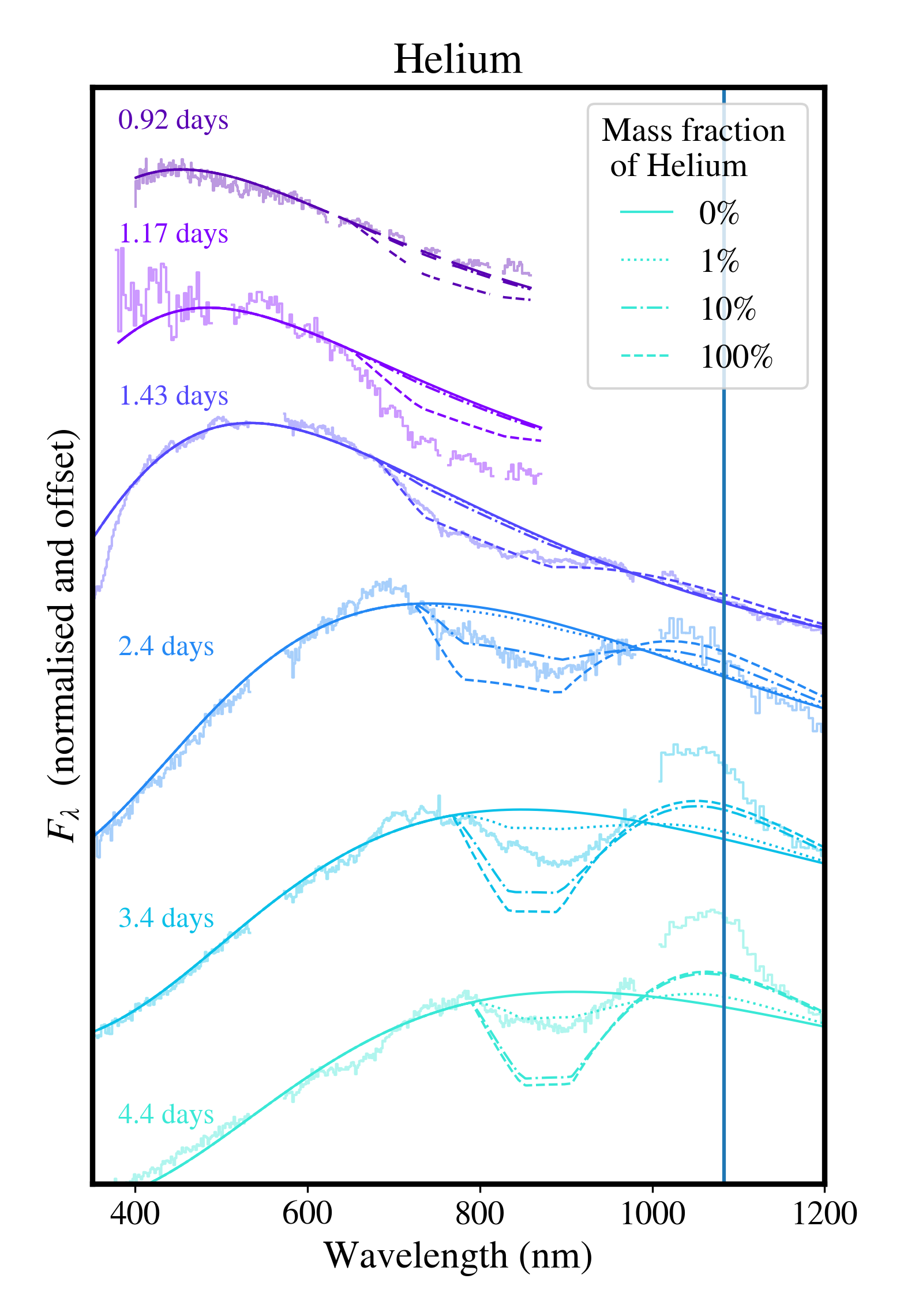} 


    \caption{Evolution of a He\,{\sc i}\,$\lambda$1083.3\,nm feature. The vertical line indicates the transition's rest wavelength. 
    In NLTE, a \hei feature would become increasingly strong with time, as the dominant pathways depleting the triplet states are from excited triplet levels that are less populated at lower temperatures. This highlights the differences in the temporal trends, which show a very weak initial but then increasingly strong He line, in contrast to the observed feature that emerges suddenly and then reduces over time.
    } 
    \label{fig:comparing_emergence}
\end{figure}

Ultimately, a \hei interpretation in early spectra is observationally contradicted in four respects. 
First, the time of the feature's first appearance is not consistent with \hei. Forming a feature at these epochs requires a helium mass comparable to the entire mass of the KN higher-velocity ejecta (i.e.\ the line-forming region). The required helium mass is lower towards the end of the photospheric epochs and begins to enter the range of nucleosynthesis calculations (see Fig.~\ref{fig:he_mass_time} and Fig.~\ref{fig:comparing_emergence}).
Second, the feature forms too suddenly for \hei. When modelling the 1\,\micron feature as \hei at 1.17\,days, such a helium-rich model would also create a feature in the earlier spectra at 0.92\,days, which is not observed. Specifically, a helium mass $M_{\mathrm{He}} > 0.004 M_{\odot}$ is ruled out at $5\sigma$ by the non-existence of a feature in the ANU spectrum from 0.92\,days, while a helium mass of $M_{\mathrm{He}} = 0.017 \pm 0.006 M_{\odot}$ is required to fit the observed feature at 1.17\,days (see Fig.~\ref{fig:comparing_emergence}). This inconsistency fundamentally follows from the large observed change in the feature, which contrasts with the relative minor evolution in radiation conditions between 0.92\,days and 1.17\,days. It is worth noting in this context that the recombination timescale, $\tau_{rec}=(n_e \alpha_{\mathrm{He\,II}})^{-1}$, can become comparable with the timescale between early spectra when given sufficiently low electron densities. For \(n_e = 10^8-10^9\,\)cm\(^{-3}\) the recombination timescale to \threeS would range from 8-0.8\,hours. This implies that i) given low enough electron-densities the steady-state approximation (i.e. using the contemporaneous radiation field) for evaluating level population may be overly simplistic and that ii) for a helium interpretation, the 1.43- to 1.47-day post-merger timescale over which the feature forms in emission requires a large electron density with $n_e \gtrsim 10^9$cm\(^{-3}\). 

Third, the higher-triplet level (\threeP) being significantly populated at early times implies the other triplet \hei lines at 587.7 and 706.7\,nm, seen in type-Ib supernovae, should also produce weaker features if a \hei 1083.3\,nm line was present. Especially, the \hei\,$\lambda 587.7$\,nm should be pronounced in the early epochs. As evaluated from the level population and $A$ values, it should have around a tenth of the strength of the 1083.3\,nm line at 1.17 and 1.43\,days. This would fade away to a few percentage points by the later epochs. This is around the typical range of line ratios seen for type~Ib supernovae \citep{Dessart2020}. A larger continuum opacity in the optical could suppress the lines, but as other features are not obscured around these wavelengths (e.g. the absorption below 400\,nm at 1.4\,days \citep{Gillanders2022,Vieira2023} and the 760\,nm P~Cygni \citep{Sneppen2023b}), such a suppression is unlikely to kill the corroborating features. Lastly, we note in passing that the 2.06\,\micron line from singlet helium (1s2s\,$^1$S--1s2p\,$^1$P) would not be observable due to the rapid radiative transition to the ground state level from 1s2p\,$^1$P, which keeps these levels depopulated \citep{Tarumi2023,Gillanders2023}.

Fourth, the observed feature strength evolves in the opposite direction expected in a helium interpretation (see Fig.~\ref{fig:comparing_emergence}). The \threeS level is increasingly easier to populate at lower temperatures (as the other triplet states become less populated and are more sensitive to photoionisation and radiative decay), so a \hei feature should grow in optical depth with time. In contrast, the observed feature decreases in optical depth. This also means the \hei mass required to produce a feature in any early epoch is inconsistent with the 1\,\micron feature for any other epoch. The helium mass required in any early epoch over-predicts the strength in all the subsequent epochs by one to two orders of magnitude, and the lower helium mass required in later epochs cannot explain the prominence of the feature observed at earlier times. We note that sizeable radial composition gradients can be produced in hydrodynamical simulations, but crucially the observed line-forming region overlaps between epochs, making the helium abundance inconsistent within the same ejecta regions. 
In Table~\ref{table:table}, we show the total ejecta mass above the photosphere (using Eq.~\ref{eq:mass}), the needed helium mass, and the corresponding helium mass fraction for each epoch. The three earliest epochs (0.92, 1.17, and 1.43\,days post merger) largely constrain the same velocity range, but they provide an order-of-magnitude inconsistency (and a 2.5$\sigma$ discrepancy) in the needed helium abundance. The subsequent daily recession of the photosphere -- at the level of 0.03 to 0.04c per day -- represents a subtle shift in the velocity-region probed when compared with the width of the line-forming region of 0.15-0.2c. While it is difficult to directly compare 1.4-day and 4.4-day constraints (given the sizeable difference in velocity range probed over longer intervals in time), day-to-day constraints are more transparently comparable, as they probe similar ejecta regions. At later times (e.g. 3.4 to 4.4\,days post merger), the required helium mass fraction displays only a subtly internal tension -- being consistent within 2$\sigma$ in derived $X_{\mathrm{He}}$. In Sect.~\ref{sec:variation_ejecta_structure}, we show that the inconsistencies in the required helium mass is not sensitive to the particular assumed radial elemental distribution but persists over a broad range of density distributions $\rho(v)$, or equivalently $\rho_{\mathrm{He}}(v)$. 

To describe the observed feature, we instead required a line that suddenly appears, remains optically thick, and then subsequently gradually becomes weaker in absorption with time. As showed by \cite{Sneppen2023_rapid, Sneppen2024}, all of these aspects are predictions of an LTE \srii interpretation of the feature. Specifically, \srii in LTE conditions has to i) become rapidly optically thick at 1.0$\pm$0.1 days in absorption, ii) remain optically thick for several days, and iii) transition to being optically thin at around three to four days post merger. Afterwards the feature should decay away. 
One could potentially argue, as mentioned in \cite{Tarumi2023}, that a blend of \srii and \hei combine to produce the observed feature, with the latter only contributing at the later epochs. However, such a blend is not observationally required by the data; the decreasing Sobolev optical depth due to the increasingly diluted \srii self-consistently captures the decreasing strength of the observed absorption. As such, the results of this temporal analysis naturally favours the conclusion found in the models of \citep{Perego2022}, where strontium rather than helium produces the observed 1\,\micron feature. 

\begin{table}[t]
\caption{ Photospheric velocity, $v_{\rm ph}$; total mass above photosphere; 
and the helium mass required above the photosphere. Lastly, given the ratio of M$_{\rm He}(>v_{\rm ph})$ to M$_{\rm ej}(>v_{\rm ph})$ one can compute the implied helium mass fraction, $X_{\rm He}$.\label{table:table} }
\renewcommand{\arraystretch}{1.4} 
\centering
\begin{tabular}{ccccc}
\hline \hline 
Time & $v_{\rm ph}$  & M$_{\rm ej}(>v_{\rm ph})$ & M$_{\rm He}(>v_{\rm ph})$ & $X_{\rm He}$ \\ \hline
[days] & [c] & [$10^{-2} M_{\odot}$] & [$10^{-2} M_{\odot}$]
\\ \hline \hline 
0.92 & $\sim$0.3 & 0.3  & $\lesssim0.3$   & $\lesssim1.0$ \\ 
1.17 & $\sim$0.3 & 0.3  & $1.2\pm0.4$     & $4 \pm 1$ \\ 
1.43 & 0.30   & 0.3  & $0.21\pm0.06$   & $0.8\pm0.3$ \\ 
2.4 & 0.27   & 0.4  & $0.05\pm0.01$ & $0.13\pm0.03$   \\ 
3.4 & 0.23   & 0.7  & $0.03\pm0.01$ & $0.05\pm0.02$  \\ 
4.4 & 0.19   & 1.0  & $0.012\pm0.005$ & $0.012\pm0.005$  \\ 
 \\ 

\hline \hline
\end{tabular} \\ \, 
\tablefoot{The photospheric velocity is ill-constrained at 0.92 and 1.17 day, as only the bluest absorption part of the 1\micron feature is within the covered spectral range. However, the line-forming region is not expected to be dramatically different between 0.92, 1.17, and 1.43 days, and the radius of the black body at these epochs supports a velocity of around 0.3c. }
\end{table}

\subsection*{Generalisations of the ejecta structure}\label{sec:variation_ejecta_structure}
The tension between the absorption of the observed 1\micron feature's evolution (i.e. fading after the first few days) and that predicted for a \hei interpretation (i.e. generically growing stronger in time) is not strongly sensitive to the specific ejecta structure modelled here. We have explored density structures \(\rho \propto v^{-\alpha}\) with various power law slopes $\alpha \in [0;10]$. Furthermore, the baseline model assumes a higher ionisation state than predicted for LTE, so we have also investigated a broader range with \(n_e = (0.6-1.2) \cdot10^8\,\)cm\(^{-3}\) (again at $v=0.3c$, $t=1$\,day), as would be suggested from single-ionised nuclei with $A\sim100$ and an ejecta mass of $0.04-0.08 M_{\odot}$ \citep[e.g.][]{Smartt2017,Cowperthwaite2017,Villar2017,Siegel2019}. Varying the electron density or helium density structure ultimately does not alleviate the inconsistency between the first days post merger and the observed feature, as this is a consequence of the diminishing strength of radiative pathways leaving triplet helium.

A fundamental limitation in the current modelling remains the assumption of spherical symmetry in the ejecta distribution. A quasi-spherical geometry is suggested by the observed spectral features in AT2017gfo \citep[see the P~Cygni features and discussions in][]{Sneppen2023,Sneppen2024}), but we note the modelling in \citet{Collins2024} shows that symmetric ejecta is not necessarily required to produce near-spherical line-forming regions from certain sight lines.
It is worth noting that the near-polar viewing angle of the merger means the absorption feature in effect only constrains the helium near the polar regions. However, fortunately for observability, it is also the polar ejecta, with its high $Y_e$-ejecta component, which is commonly interpreted as allowing the highest helium abundance \citep[e.g.][]{Kawaguchi2022,Just2023}.
Constraints on the equatorial ejecta could be obtained from studying the emission component, but such examinations require modelling of reverberation effects \citep[e.g.][]{Sneppen2023_rapid} and assumptions on the underlying radiative field (which for these distant ejecta parts are not directly observed or known), thus making any conclusions more difficult to assess. Full 3D radiative transfer modelling on numerical relativity simulated KNe (such as work in \citet{Shingles2023,Collins2024}) are a natural next step in exploring the robustness of the constraints presented here, but we emphasise even large geometric perturbations are unlikely to alleviate the fundamental inconsistency in the helium model's evolution, namely that the observed declining strength of the radiative field results in an order-of-magnitude less suppression of a \hei triplet population. 

\section{Assessing the importance of NLTE modelling for kilonovae at early times?}\label{sec:NLTE_required}

Perhaps the most striking aspect of the early spectra taken of AT2017gfo is their apparent simplicity. A simple black body empirically describes the continuum very well at early times \citep[e.g.\ the first X-shooter spectrum at 1.4\,days has percent-level consistency in the inferred temperature from the UV through the NIR, see][]{Sneppen2023_bb}. The spectral perturbations away from the black body are produced in the KN atmosphere by the \rprocess species with the strongest individual transitions, with LTE appearing to be a good assumption for the relevant ionisation state of those species \citep{Watson2019,Domoto2022}. Indeed, the emergence time and early spectral shape of features in AT2017gfo follow the predictions of a recombination wave between ionisation states in LTE, which constrains the black-body temperature of the radiation field to be within 5\% of the electron temperature \citep{Sneppen2024}. While both the continuum and the spectral features are thus well-described by LTE conditions and LTE populations at early times, it remains unclear as to what extent NLTE effects are required and, indeed, permitted by the early data.

We note further complexities could be explored in future models, such as the variety of ejecta structures, including more refined radioactive ionisation modelling, and allowing different excitation temperatures beyond the observed colour temperature. However, as noted above, the NLTE model developed by \citet{Tarumi2023} and extended here already assumes optimistic conditions for forming a He feature. That it still cannot describe the observed feature is because one would need to produce an NLTE model that populates the \threeS level significantly without populating any other excited levels of \hei. This seems difficult since one would have to decouple \threeS from \threeP, and these are respectively the relevant lower and upper levels of the 1083\,nm transition. 

One cannot alleviate the non-physical helium mass required at early times by introducing stronger (or weaker) radioactive ionisation because the models presented already assume the majority of helium is singly ionised and thus already assume the maximum recombination rate possible. Indeed, any significant change to the radioactive ionisation modelling would only decrease \heii, thus decreasing the recombination rate to \hei and increasing the helium mass tension. %
\citet{Tarumi2023} note that non-thermal electrons should deplete \srii within their framework in favour of more highly ionised species with time. Conversely, the NLTE modelling in \cite{Pognan2023} still finds substantial contribution from \srii, even for late-time spectra. Ultimately, current ionisation state modelling is highly uncertain for \rprocess elements. Due to the lack of photoionisation cross-sections for \rprocess elements, the recombination rates are simply scaled hydrogenic approximations in \citet{Tarumi2023}, and non-thermal collisions are commonly approximated with generic formulae for ionisation cross-sections \citep[see][]{Pognan2022}. Improvements to the atomic data through detailed theoretical calculations and experiments will be instrumental to permitting stronger statements based on the observed data.

Nevertheless, it is worth considering what the observed consistency of the 1\,\micron feature (and its continued relation to \srii) suggests for modelling radioactive ionisation because models that predict that the \srii population should substantially diminish in time (due to radioactive ionisation) do not describe the observed feature's evolution well. For context, we note that detonation models of type Ia supernovae predict that non-thermal ionisation leads to the presence of multiply ionised species and the depletion of low-ionisation species, which again is in contrast to the observed emission lines from Type\,Ia supernova, which are dominated by singly  and doubly ionised Fe-group nuclei \citep{Wilk2018,Shingles2022}. Such discrepancies may potentially be alleviated by introducing clumping \citep{Wilk2020}, while it may also hint at broader tensions between current non-thermal ionisation modelling and the observed spectra.

\section{Conclusion}
In this analysis, we have revisited whether \hei\,$\lambda 1083.3$\,nm in NLTE can explain the observed \(1\,\mu\)m P~Cygni feature in the KN AT2017gfo. Employing a collisional-radiative model, we found that the density of triplet helium (and thus any \hei\,$\lambda 1083.3$\,nm feature) is highly temperature sensitive. Due to the decreasing importance of radiative transitions, a helium feature will generically become stronger in time, which is in contrast to the decreasing Sobolev optical depth of the observed feature. Thus, the helium mass required to produce the \(1\,\mu\)m feature is inconsistent between each observed epoch.

The earliest appearance of the feature is in the SALT spectrum at 1.17\,days, where the helium mass required in the line-forming region, $M_{\rm He}\sim10^{-2}\,{\rm M}_{\odot}$, is comparable to (or in excess of) the entire high-velocity ejecta component (i.e.\ $X_{\rm He} \sim 1$), which effectively excludes a \hei interpretation from a nucleosynthesis perspective at early times on mass grounds alone. This interpretation is further disfavoured by the non-detection of corroborating features (particularly \hei\,$\lambda 587.7$\,nm), the emergence time being more rapid than suggested by the feature's formation, and the subsequent fading in the line.
While the \srii model, under the assumption of LTE, can reproduce the data at all times, the increased strength of a \hei line with time implies a potential contribution is possible at later epochs for nucleosynthesis models where sufficient helium is produced. Indeed, observations in these epochs may be used to constrain the maximum helium abundance allowed in the kilonova AT2017gfo \citep[as detailed in][]{Sneppen2024_lifetime}. 
Aggregated atomic data for the \rprocess elements is needed to help understand the constraints the formation, existence, and evolution of the observed spectral features place on the ejecta.




\section*{Data availability}
The spectral series used in this paper is composed from a series of different observing programmes at various telescopes. The 0.92\,day spectra were obtained with the Australian National University (ANU) the 2.3\,meter telescope located at Siding Spring Observatory. The 1.17\,day data obtained with the Southern African Large Telescope (SALT) under the Director’s Discretionary Time programme 2017-1-DDT-009, are available at \url{https://ssda.saao.ac.za} with the newly reduced spectra (e.g.\ with improved flux calibration, see \cite{Sneppen2024}) now publicly available at \url{https://github.com/Sneppen/Kilonova-analysis}. X-shooter data from European Space Observatory (ESO) telescopes at the Paranal Observatory under programmes 099.D-0382 (principal investigator [PI]: E.~Pian), 099.D-0622 (PI: P.~D’Avanzo), 099.D-0376 (PI: S.~J.~Smartt), which are available at \url{http://archive.eso.org} and WISeREP (\url{https://wiserep.weizmann.ac.il/}). The compiled data-sets (and re-processed when necessary, see \cite{Sneppen2024}) can be found at \url{https://github.com/Sneppen/Kilonova-analysis}.

For P~Cygni profiles, we use the implementation in the Elementary Supernova \citep{Jeffery1990} from \url{https://github.com/unoebauer/public-astro-tools} with generalisations for computational speed-up and to account for special relativistic corrections as outlined in \cite{Sneppen2023A&A} see \url{https://github.com/Sneppen/Kilonova-analysis}. The code to estimate the NLTE level populations for helium is provided in \url{https://github.com/rasmus98/NLTE-Helium}.

\begin{acknowledgements}
The authors would like to thank Kenta Hotekezaka for clarifying and discussing the NLTE helium modelling in previous literature. Additionally, the authors would like to thank Oliver Just, Gabriel Martínez-Pinedo, Andreas Bauswein, Albino Perego, Leonardo Chiesa, Nick Vieira and Zewei Xiong for discussing the yields and variations within helium nucleosynthesis modelling. 

The Cosmic Dawn Center (DAWN) is funded by the Danish National Research Foundation under grant DNRF140. AS, RD, DW and SAS are funded/co-funded by the European Union (ERC, HEAVYMETAL, 101071865). Views and opinions expressed are, however, those of the authors only and do not necessarily reflect those of the European Union or the European Research Council. Neither the European Union nor the granting authority can be held responsible for them. LJS acknowledges support by the European Research Council (ERC) under the European Union’s Horizon 2020 research and innovation program (ERC Advanced Grant KILONOVA No. 885281). LJS acknowledges support by Deutsche Forschungsgemeinschaft (DFG, German Research Foundation) - Project-ID 279384907 - SFB 1245 and MA 4248/3-1. CEC is funded by the European Union’s Horizon Europe research and innovation programme under the Marie Skłodowska-Curie grant agreement No. 101152610.

\end{acknowledgements}





\bibliographystyle{mnras}
\bibliography{refs} 

\setcounter{section}{1}
\setcounter{equation}{0}
\setcounter{figure}{0}
\renewcommand{\thesection}{Appendix \arabic{section}}
\renewcommand{\theequation}{A.\arabic{equation}}
\renewcommand{\thefigure}{A.\arabic{figure}}


\end{document}